# CNT effective interfacial energy and pre-exponential kinetic factor from measured NaCl crystal nucleation time distributions in contracting microdroplets

Ruel Cedeno[1,2], Romain Grossier[1]*, Nadine Candoni[1], Nicolas Levernier[3], Adrian Flood[2]*, Stéphane Veesler[1]*

[1]CNRS, Aix-Marseille University, CINaM (Centre Interdisciplinaire de Nanosciences de Marseille), Campus de Luminy, Case 913, F-13288 Marseille Cedex 09, France

[2]Department of Chemical and Biomolecular Engineering, School of Energy Science and Engineering, Vidyasirimedhi Institute of Science and Technology, Rayong 21210, Thailand

[3]INMED, INSERM, Aix Marseille Univ, France, Turing Centre for Living systems, Marseille, France ; CPT : Aix Marseille Univ, Université de Toulon, CNRS, CPT (UMR 7332), Turing Centre for Living systems, Marseille, France

***Authors to whom correspondence should be addressed**: adrian.flood@vistec.ac.th; romain.grossier@cnrs.fr; stephane.veesler@cnrs.fr

## ABSTRACT

Nucleation, the birth of a stable cluster from disorder, is inherently stochastic. Yet up to date, there are no quantitative studies on NaCl nucleation that accounts for its stochastic nature. Here, we report the first stochastic treatment of NaCl-water nucleation kinetics. Using a recently developed microfluidic system and evaporation model, our measured interfacial energies extracted from a modified Poisson distribution of nucleation time show an excellent agreement with theoretical predictions. Furthermore, analysis of nucleation parameters in 0.5 pL, 1.5pL and 5.5 pL microdroplets reveals an interesting interplay between confinement effects and shifting of nucleation mechanisms. Overall, our findings highlight the need to treat nucleation stochastically rather than deterministically to bridge the gap between theory and experiment.

## I. INTRODUCTION

Nucleation plays a key role in a wide array of applications including nanosynthesis[1], energy storage[2], pharmaceutical production, biomineralization, and climate modeling.[3] Sodium chloride, being the most abundant salt on earth[4], is of particular interest due to its influence on metal corrosion[5], building material degradation[6], oil well productivity[7], atmospheric science[3] and so on. Thus, fundamental understanding of its nucleation kinetics is of paramount importance yet it remains poorly understood from both experimental and theoretical perspective.[8] Indeed, nucleation rates from simulations still differ from experiments by several orders of magnitude.[8, 9] With diverse simulation approaches being applied,[8, 10-14] reliable benchmarking of computational results with experiments remains a challenge as there are only very few experimental studies that quantitatively measure the nucleation kinetic parameters of NaCl in water. These include experiments which employ an efflorescence chamber[15], electrodynamic levitator[16], and microcapillaries[4]; all of which treated nucleation deterministically. In this context, deterministic methods calculate nucleation rates directly from the mean nucleation time whereas stochastic methods employ the probability distribution of nucleation times. However, primary nucleation has been shown to be inherently stochastic rather than deterministic.[17, 18] Moreover, because nucleation is a rare event and the critical nucleus is a transient species, it is barely undetectable by classical experiment. This can be addressed by introducing a local bias in the solution so that nucleation is much more probable.[19] Recently, with *in situ* electron microscopy, Nakamuro et al.[20] have captured atomically-resolved images of NaCl nucleation in confined conical carbon nanotubes (with volume confinement as a bias). They observed that a critical cluster must have at least 48 NaCl units and that the nucleation periods follow a normal distribution spanning from 2 to 10 s based on nine nucleation events. Although more data points are needed to generate a reliable statistical distribution, this is a strong evidence for the stochasticity of NaCl nucleation yet surprisingly, there are no existing experimental studies that measure its kinetic parameters using the stochastic view of nucleation. In this communication, we address this by measuring the primary nucleation kinetic parameters of aqueous NaCl in confined microdroplets, in the pL range, with a stochastic model. We demonstrate that by combining the deliquescence-recrystallization cycle for measuring induction times[21], an appropriate evaporation model[22], together with an inhomogeneous Poisson probability distribution of nucleation time[23] and Classical Nucleation Theory (CNT), one can obtain a reliable estimate of nucleation parameters, effective interfacial energy[24] and pre-exponential kinetic factor, which are consistent with theoretical and experimental values from literature. Finally, we analyze the impact of confinement by volume[25,26] on the measured kinetic parameters.

## II.    MATERIALS AND METHODS

Arrays of sessile NaCl microdroplets are generated on PMMA-coated glass immersed in a thin layer of PDMS oil at ambient conditions (1 atm, 25°C). Supersaturation to achieve nucleation is obtained via the droplet contraction technique[27] (similar to an evaporation process) under controlled humidity (10% *RH*) using the method described in our previous work.[21] A schematic overview of the setup and illustration of the droplet contraction technique is shown in **Figure 1a-b** while selected microdroplet images are shown in **Figure 1c-d**.

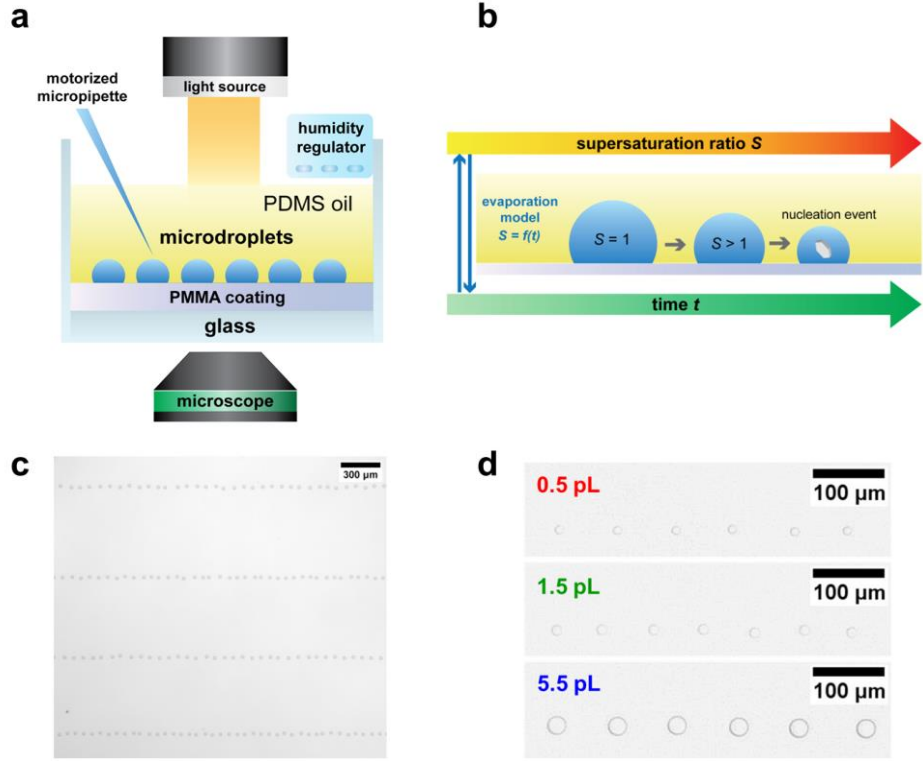

**Figure 1**(a) Schematic overview of the experimental setup (adapted from Ref[22]) (b) illustration of droplet evaporation until nucleation (c) image of a typical 2-D array of microdroplets (d) image of microdroplets at saturation across different sizes at saturation.

For each droplet, we calculate the dimensionless nucleation time[21] τ i.e. the time elapsed between saturation and the nucleation point divided by a characteristic time accounting for the evaporation rate.[21] This is a pre-processing procedure to robustly take into account the variability of evaporation rate in the set-up.[28] We also verified that there is no interference of diffusion-mediated interactions between microdroplets.[21] To find the volume and supersaturation ratio as a function of τ, we employed an evaporation model tailored to our specific configuration.[22] We then plot the cumulative probability distribution of supersaturation ratio at nucleation (i.e. fraction of the microdroplets that has nucleated at a given supersaturation ratio) for each set of microdroplet sizes. We then fit distribution of nucleation times with the help of an inhomogeneous Poisson distribution, as proposed by Goh et al.[29], under conditions of time-varying supersaturation

$$P(t) = 1 - \exp\left[-\int_{t_{\text{sat}}}^{t_{\text{nuc}}} J(t)V(t)dt\right] \quad (1)$$

with *P(t)* as the fraction of microdroplets nucleated after time *t*, *V(t)* droplet volume at time *t*, $t_{sat}$ time at which the microdroplet becomes saturated and $t_{nuc}$ nucleation time. For the primary nucleation rate *J(t)*, we used the classical nucleation theory (CNT) as

$$J(t) = A \exp\left[-\frac{16\pi}{3}\frac{\gamma_{eff}^3}{\rho_s^2(k_bT)^3 \ln^2 S(t)}\right] \quad (2)$$

where $A$ is the pre-exponential factor, $\gamma_{\text{eff}}$ is the effective interfacial energy, $\rho_s$ is the number density of formula units in the solid ($2.27 \times 10^{28}$ m$^{-3}$ for NaCl), $k_bT$ is the thermal energy, and $S(t)$ is the supersaturation expressed as the ratio of concentrations (concentration at nucleation / concentration at saturation) at nucleation at time $t$. This represents a general equation for primary nucleation with $\gamma_{\text{eff}}$ accounting for the "degree" of heterogeneity.

### III. RESULTS AND DISCUSSION

With the nucleation times obtained via image analysis, we computed the normalized nucleation time $\tau$ and the results are plotted in **Figure 2a.** Using our tailored evaporation model, $\tau$ can then be converted into distributions of supersaturation ratios at nucleation, as presented in **Figure 2b**. As **Figure 2b** shows, the combination of the inhomogeneous Poisson distribution (eq 1) and classical nucleation theory (eq 2) well captures the sigmoidal nature of the distribution. Unlike the use of empirical distributions (such as Weibull, Gompertz, Gumbell etc), whose parameters cannot be interpreted in terms of CNT, this method allows the extraction of the kinetic parameter $A$ (pre-exponential factor) and thermodynamic parameter $\gamma_{\text{eff}}$ (effective interfacial energy between crystal and solution). In theory, $\gamma_{\text{eff}}$ reflects how the thermodynamic barrier is reduced due to heterogeneous nucleation, that is, if $\gamma_{\text{HOM}}$ is the homogeneous interfacial energy, then $\gamma_{\text{eff}}$ lies within $0 < \gamma_{\text{eff}} < \gamma_{\text{HOM}}$. The fitted nucleation parameters are listed in **Table 1**. These parameters can then be used to calculate the nucleation rate $J$ and apparent critical size $n^*$ as a function of supersaturation ratio $S$ which are shown in **Figure 2c** and **Figure 2d** respectively. Note that the apparent critical size $n^*$ is calculated based on an equivalent spherical nucleus with an interfacial energy of $\gamma_{\text{eff}}$. This is an approximation to the actual critical size which theoretically depends on the contact angle between the nucleus and the substrate[24] as well as the shape factor of the nucleus.[30] Moreover, since $n^*$ is calculated based on $\gamma_{\text{eff}}$, it also takes into account the effect of heterogeneous nucleation.

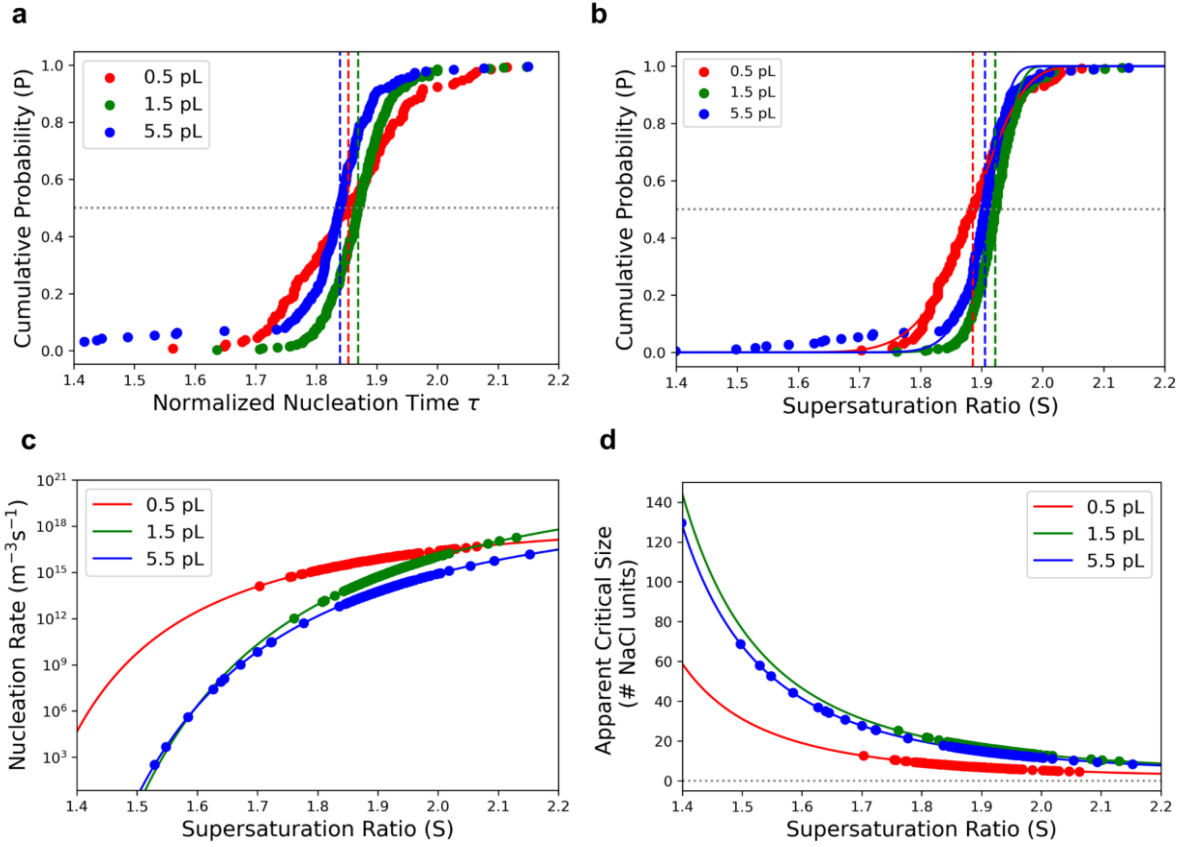

**Figure 2.** (a) Cumulative distribution of Normalized Nucleation Time τ. The median τ is marked by dashed vertical lines and $P = 0.5$ is marked by grey dotted line. (b) Fitting of the inhomogeneous Poisson distribution (Equation 1) with the experimental distribution of supersaturation ratio at nucleation. The median $S$ is marked by dashed vertical lines. (c) nucleation rate (J in $m^{-3}s^{-1}$) as a function of supersaturation ratio computed using the kinetic parameters in Table 1. (d) Apparent critical size (# of NaCl units) as a function of supersaturation ratio computed using the fitted effective interfacial energy $\gamma_{eff}$.

**Table 1** Nucleation kinetic parameters and their corresponding standard error obtained from the fit in Figure 2b.

| Volume at saturation $V_{sat}$ (pL) | Median $S$ at nucleation, $\bar{S}_n$ | Effective Interfacial Energy $\gamma_{eff}$ (mJ/m$^2$) | Kinetic Prefactor $\log_{10} A (m^{-3}s^{-1})$ |
|---|---|---|---|
| 0.5 | 1.89(±3%) | 47.5 (±0.4%) | 19.8 (±0.2%) |
| 1.5 | 1.92(±3%) | 64.5 (±0.2%) | 24.4 (±0.3%) |
| 5.5 | 1.91(±5%) | 61.9 (±0.4%) | 22.3 (±0.5%) |

To explain the observed trends with respect to volume in Figure 2 and Table 1, we consider two relevant phenomena: (1) the interplay between homogeneous and heterogeneous mechanism (2) confinement effects (kinetic and thermodynamic).

Although the median supersaturation $\bar{S}_n$ are essentially identical for the three studied volumes, the effective interfacial energy $\gamma_{eff}$ of the 0.5pL set is significantly lower than that of the rest.

Note that a lower effective thermodynamic barrier is characteristic of heterogeneous nucleation mechanism. As smaller droplets have higher surface area to volume ratio, the probability of surface nucleation becomes more apparent relative to bulk nucleation. Moreover, lower supersaturation generally favors heterogeneous mechanism while higher supersaturation $S$ generally favors homogeneous nucleation. This is consistent with the plot of nucleation rate $J$ against supersaturation ratio (**Figure 2c**) where we see that at lower $S$, the smallest droplet size nucleates faster. However, if the process is governed entirely by the interplay between heterogeneous and homogeneous mechanism, then we would expect that the thermodynamic barrier $\gamma_{\text{eff}}$ should obey 0.5 pL < 1.5 pL < 5.5 pL. Interestingly, what we observe for $\gamma_{\text{eff}}$ is 0.5 pL < 5.5 pL ≤ 1.5 pL. This suggests that other phenomena must be at play. Since the droplet volume is in picoliter range, kinetic and thermodynamical confinements could have an impact. Note that kinetic confinement stems from the fact that nucleation time scales inversely with the nucleation rate $J$ and system volume $V$, i.e. $t_n \propto \frac{1}{JV}$. On the other hand, thermodynamic confinement originates from the depletion of the effective supersaturation[24] level during the formation of the pre-critical clusters in finite-sized system. In such finite systems, critical size is determined at a lower "effective" supersaturation, so the critical size would be larger than that of an infinite systems where no depletion occurs. In other words, translated in an infinite system where such confinement through depletion is not taken into account (as in the proposed model herein), it would correspond to a (virtually) higher effective surface energy $\gamma_{\text{eff}}$. However, the exact quantification of such thermodynamic confinement through depletion effects as sub-critical population emerge in solution is difficult. This would necessitate, at least, the exact knowledge of the distribution of pre-critical clusters. Nevertheless, we could get some trends from previous works: the higher the solute solubility, the lesser the impact of thermodynamic confinement. For instance, if one compares NaCl and AgCl (known for its low solubility in water) using a "toy model" [24], we see that the depletion effects in supersaturation needed to form a single critical-cluster has a minor role in the case of NaCl, but a dramatic effect in the case of AgCl (as sketched in Figure S5 of SI). This could explain the slight increase in $\gamma_{\text{eff}}$ as the volume decreases from 5.5 pL to 1.5 pL. In other words, the observed trend in $\gamma_{\text{eff}}$ could originate from the competition between heterogeneous nucleation and confinement effects, i.e. heterogeneous nucleation tends to decrease $\gamma_{\text{eff}}$ while confinement tends to increase $\gamma_{\text{eff}}$. Specific experiments to address this particular problem would need to be made, with a larger bandwidth in addressed droplet volumes and comparison with salts/molecules of diverse solubilities. Regarding the kinetic prefactor, its interpretation is generally more complex as it is related to the mass transfer rate of the monomers to the cluster surface which is a function of the attachment-detachment frequency, viscosity, diffusivity, Zeldovich factor, concentration of nucleation sites etc.

Nevertheless, we remind the reader that the analysis presented here is based on classical nucleation theory (CNT), which has been shown to have inherent limitations particularly at high supersaturations. In such conditions, critical size shrinks to dramatically small sizes where the discontinuity of matter (atoms) should play a role, or at least be the source of large deviations from CNT set of hypothesis (in particular the capillary approximation[31]). Although non-classical nucleation theories have been developed[32, 33], their governing equations are more complex and contain multiple fitting parameters. For this reason, we chose CNT to model our experimental data. Moreover, using CNT would allow us to compare our results against the literature (both experiments and simulations). For instance, our fitted $\gamma_{eff}$ ranges from 45.3 to 61.1 mJ/m² (Table 1). These values are consistent with that of Hwang et al.[34] who reported $\gamma_{eff}$ = 46.17 mJ/m² via electrostatic levitation experiment. Interestingly, a seeded atomistic simulations performed by the group of Peters[10] resulted in an interfacial energy of $\gamma$ = 47 mJ/m² while the molecular dynamics simulation of Bahadur et al[12] yielded a $\gamma$ of 63 mJ/m. This is a reasonable agreement between theory and experiment[2].

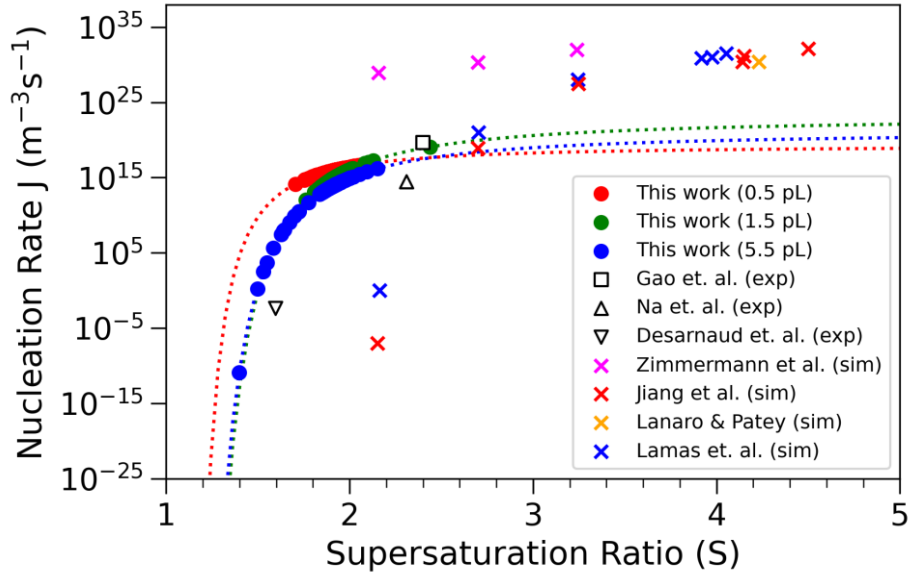

**Figure 3.** Nucleation rate J, of NaCl in water as a function of supersaturation. Comparison of our experimental results (dashed blue line is via extrapolation of CNT) to relevant experimental literature data (exp) and theoretical simulations (sim). Experiments were based on efflorescence chamber (Gao et. al.)[15], spherical void electrodynamic levitator trap (Na. et. al.)[16], and microcapillaries (Desarnaud et. al.)[4] while the simulations were based on seeded atomistic simulations (Zimmerman et. al.)[8, 10], forward flux sampling (Jiang et. al.)[14], direct molecular dynamics (Lanaro & Patey)[13], and seeding simulations (Lamas et al.)[35].

For further comparison with the literature, we then plotted the nucleation rate *J* as a function of supersaturation ratio *S* in **Figure 3** together with the experimental and simulated data from multiple research groups. We can see that the magnitude of our measured nucleation rate is very close to that of Gao et. al[15] measured in an efflorescence chamber experiment and Na. et. al[16] who used an electrodynamic levitator trap, a setup that aimed to minimize all possible heterogeneous nucleation sites. Although they reported an interfacial energy between crystal and solution $\gamma$ = 87 mJ/m², they calculated it from the average induction time while taking $A$ = $10^{30}$ m⁻³s⁻¹ as a fixed value (taking induction time as deterministic rather than stochastic). Interestingly, when we used a similar calculation procedure (average induction time and $A$ = $10^{30}$ m⁻³s⁻¹), we obtained a value of effective interfacial energy $\gamma_{eff}$ of 77, 79, and 80 mJ/m² for 0.5 pL, 1.5 pL, and 5.5 pL respectively.

Thus, the discrepancy in the measured interfacial energy is likely due to two main reasons. First, their approach assumes nucleation as a deterministic process (based on average induction time) while our treatment considers its inherent probabilistic nature (a more realistic view of nucleation). Second, we did not assume any pre-defined value of the pre-exponential factor in the parameter estimation. In the experimental work of Gao et. al[15] where they measured mean efflorescence time, they also fixed the prefactor at a value of $2.8 \times 10^{38}$ $m^{-3}s^{-1}$. Furthermore, in the microcapillary experiments of Desarnaud et. al.[4], they reported $J = 0.004$ $m^{-3}s^{-1}$ at $S \approx 1.6$ but they fixed the value of $\gamma$ at 80 mJ/m$^2$. Thus, to the best of our knowledge, our work is the first experimental work that employed a probabilistic approach to measure the interfacial energy between crystal and solution for NaCl-water system without assuming a fixed value of pre-exponential factor. This suggests that the commonly accepted experimental value of $A$ and $\gamma$ for NaCl crystallization may need to be re-examined. Given that the current theoretical simulations generally overestimate the experimental nucleation rates (**Figure 2b**), our findings can serve as an additional benchmark leading to new insights which could bridge the gap between theory and experiments.

Overall, we highlight that these interesting finite-size effects are clearly observable in our microfluidic experiments which would not be observed in bulk solution experiments. The data treatment of our experiments with CNT model allows us to have a better understanding of nucleation, providing kinetic and thermodynamic information on the system NaCl/water.

## IV. CONCLUSION

To summarize, we report a stochastic approach to extract the nucleation kinetic parameters from the induction time distribution of evaporating sessile microdroplets, using NaCl-water as a model system. We showed that by combining a modified Poisson distribution analysis together with an accurate evaporation model, one could obtain reliable nucleation kinetic parameters (both kinetic and thermodynamic). Our results also reveal the competition between the nucleation-enhancing heterogeneous mechanism and the nucleation-inhibiting confinement effects. However, to fully elucidate the underlying mechanisms, it would be interesting to investigate a wider range of droplet sizes together with finer control of evaporation rate. To investigate quantitatively the impact of thermodynamic confinement, modeling the distributions of the pre-critical clusters would be essential. The use of non-classical nucleation theories can also be explored.

Given the numerous simulation studies on NaCl nucleation[8, 10-14], our experimental nucleation parameters based on stochastic approach presented here can serve as an additional benchmark in validating theoretical predictions. Moreover, our experimental approach and data-treatment protocol can also be extended to study the nucleation of other salts, biological, and pharmaceutical crystals of interest.

# SUPPLEMENTARY MATERIAL

Classical Nucleation Theory for Ionic Systems
Modified Poisson Distribution Function
Ionic Activity Coefficient vs Supersaturation Ratio
Evaporation Model
Data Processing and Curve Fitting Method
Effect of Solubility on Thermodynamic Confinement

# AUTHOR'S CONTRIBUTIONS

**Ruel Cedeno** : Formal analysis (equal); Methodology (equal); Writing − review & editing (equal).

**Romain Grossier** : Supervision (equal); Formal analysis (equal); Conceptualization (equal); Writing − review & editing (equal). **Nicolas Levernier** : Conceptualization (supporting). **Nadine Candoni** : Supervision (equal); Conceptualization (equal);. **Adrian Flood** : Supervision (equal); Writing − original draft (equal). **Stéphane Veesler** : Supervision (equal); Conceptualization (equal); Writing − original draft (equal);

# ACKNOWLEDGMENT

RC acknowledges the financial support of Vidyasirimedhi Institute of Science and Technology (VISTEC) and the Eiffel Excellence Scholarship (N°P744524E) granted by the French Government. NL was supported by the French National Research Agency (ANR-16-CONV-0001) and from Excellence Initiative of Aix-Marseille University - A*MIDEX.

# AUTHOR INFORMATION

**Ruel Cedeno** - CNRS, Aix-Marseille University, CINaM (Centre Interdisciplinaire de Nanosciences de Marseille), Campus de Luminy, Case 913, F-13288 Marseille Cedex 09, France; Department of Chemical and Biomolecular Engineering, School of Energy Science and Engineering, Vidyasirimedhi Institute of Science and Technology, Rayong 21210, Thailand;
ORCID: https://orcid.org/0000-0001-9948-3943

**Romain Grossier** - CNRS, Aix-Marseille University, CINaM (Centre Interdisciplinaire de Nanosciences de Marseille), Campus de Luminy, Case 913, F-13288 Marseille Cedex 09, France;
ORCID : https://orcid.org/0000-0002-5207-2087; Email : romain.grossier@cnrs.fr

**Nicolas Levernier -** INMED, INSERM, Aix Marseille Univ, France, Turing Centre for Living systems, Marseille, France ; CPT : Aix Marseille Univ, Université de Toulon, CNRS, CPT (UMR 7332), Turing Centre for Living systems, Marseille, France

**Nadine Candoni** - CNRS, Aix-Marseille University, CINaM (Centre Interdisciplinaire de Nanosciences de Marseille), Campus de Luminy, Case 913, F-13288 Marseille Cedex 09, France
ORCID : https://orcid.org/0000-0002-7916-7924

**Adrian Flood** - Department of Chemical and Biomolecular Engineering, School of Energy Science and Engineering, Vidyasirimedhi Institute of Science and Technology, Rayong 21210, Thailand;
ORCID: https://orcid.org/0000-0003-1691-3085 ; Email: adrian.flood@vistec.ac.th

**Stéphane Veesler** - CNRS, Aix-Marseille University, CINaM (Centre Interdisciplinaire de Nanosciences de Marseille), Campus de Luminy, Case 913, F-13288 Marseille Cedex 09, France; ORCID: https://orcid.org/0000-0001-8362-2531  ; Email: stephane.veesler@cnrs.fr

**Supplementary Material for:**

# CNT effective interfacial energy and pre-exponential kinetic factor from measured NaCl crystal nucleation time distributions in contracting microdroplets

Ruel Cedeno[1,2], Romain Grossier[1], Nadine Candoni[1], Nicolas Levernier[3], Adrian Flood[2]*, Stéphane Veesler[1]*

[1]CNRS, Aix-Marseille University, CINaM (Centre Interdisciplinaire de Nanosciences de Marseille), Campus de Luminy, Case 913, F-13288 Marseille Cedex 09, France

[2]Department of Chemical and Biomolecular Engineering, School of Energy Science and Engineering, Vidyasirimedhi Institute of Science and Technology, Rayong 21210, Thailand

[3]INMED, INSERM, Aix Marseille Univ, France, Turing Centre for Living systems, Marseille, France ; CPT : Aix Marseille Univ, Université de Toulon, CNRS, CPT (UMR 7332), Turing Centre for Living systems, Marseille, France

**S1. Classical Nucleation Theory for Ionic Systems**

Classical Nucleation theory expresses the primary nucleation rate *J* as the product of the pre-exponential factor *A* and an exponential factor containing the free energy cost of forming a critical nucleus *ΔG\** and thermal energy $k_bT$.

$$J = A \exp\left(-\frac{\Delta G^*}{k_b T}\right) \tag{S1}$$

An important difference between the treatment of ionic systems and molecular systems is in the expression of chemical potential difference between solid and liquid.[1] For ionic systems, it is a function of the number of ions forming one formula unit v ( 2 for NaCl), and the mean ionic activity coefficient of the solute $\gamma_\pm$. These leads to the following expression for ΔG*

$$\Delta G* = \frac{4}{3}\pi\gamma(R_c)^2 \quad \text{with} \quad R_c = \frac{2\gamma}{vkT\rho_s \ln\left(\frac{\gamma_\pm}{\gamma_{\pm_0}}S\right)} \tag{S2}$$

with interfacial energy γ between crystal and solution, critical radius $R_c$, number density of formula units in the solid $\rho_s$ (2.27 × 10$^{28}$ m$^{-3}$ for NaCl), and *S* is the supersaturation ratio (c/c$_{sat}$)[1].

## S2. Modified Poisson Distribution Function

In the stochastic view of nucleation, the probability distribution of the nucleation times must be analyzed. In the context of microdroplets, it is normally assumed that the time it takes for a nucleus to grow to detectable size is negligible[2].

Thus, for constant supersaturation experiments, the cumulative probability of obtaining a droplet with at least one nucleus after time $t$ is a function of nucleation rate $J$ and droplet volume $V$ given as

$$P(t) = 1 - \exp(-JVt) \tag{S3}$$

In the case of evaporating droplet, both the supersaturation and the volume vary with time. As suggested by Goh et. al.,[2] the cumulative probability distribution function becomes

$$P(t) = 1 - \exp\left[-\int_{t_{sat}}^{t_{nuc}} J(t)V(t)dt\right] \tag{S4}$$

In equation (S4), $J(t)$ can be expressed as a function of supersaturation $S(t)$ by combining with equations (S1) through (S2). The value of $V(t)$ and $S(t)$ was obtained from an evaporation model (section S4).

## S3. Ionic Activity Coefficient vs Supersaturation Ratio

The ionic activity coefficient is necessary to calculate the chemical potential from concentration. To obtain the activity coefficient as a function of supersaturation ratio, we employed the experimental data of Na et al.[3] We then used an empirical function (logistic) to fit the data.

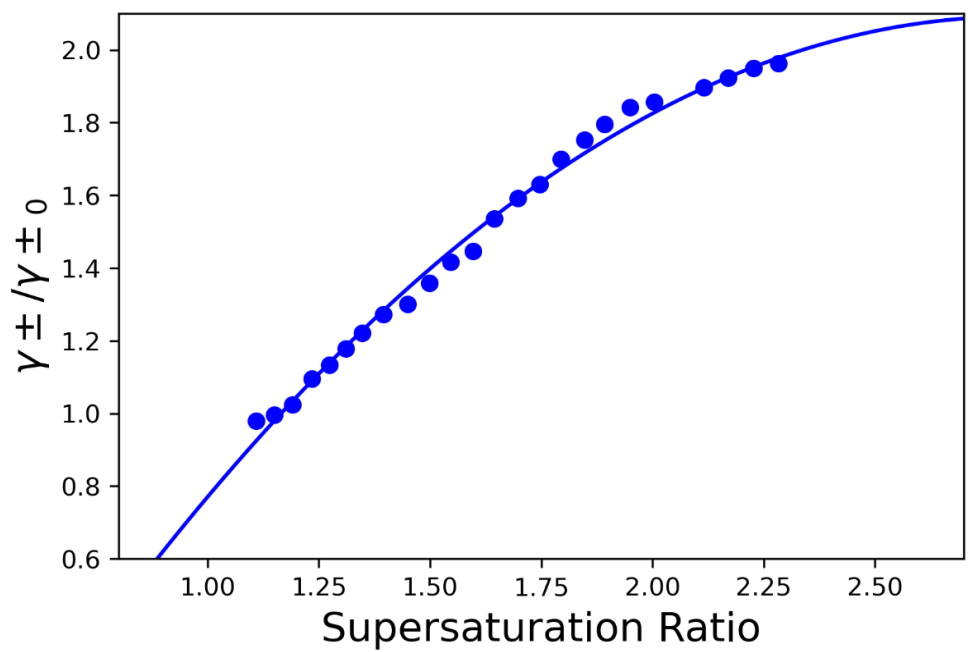

**Figure S1.** Ratio of ionic activity coefficients $\gamma_\pm/\gamma_{\pm 0}$ based on the experimental data of Na et al.[3] The data is fitted with a general logistic function $y = a/(1+\exp(-b(x-c)))$ resulting in $a$ = 2.1889, $b$=2.147, $c$=1.241 with $R^2$ = 0.995.

## S4. Evaporation Model

### S4.1. Comparison with constant evaporation rate model

To determine the supersaturation ratio as a function of nucleation time, we used an evaporation model tailored for our specific system.[4] The evolution of volume and supersaturation with time are plotted as follows. As expected, smaller droplets evaporate faster due to higher surface area to volume ratio. For comparison, the prediction of a simple linear model (constant evaporation rate) is shown (dotted line).

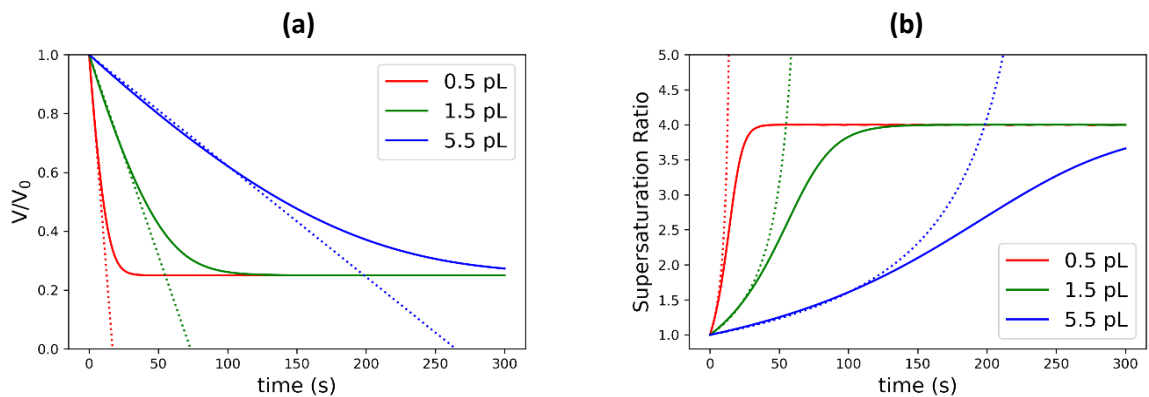

**Figure S2**. (a) Time evolution of normalized volume ($V/V_0$) and (b) supersaturation ratio, calculated using our tailor-made evaporation model (solid line)[4] and a simple linear model (dotted line).

The use of simple linear model has been shown to overestimate the supersaturation ratio at nucleation particularly at the later stages of the evaporation process where the changes in water activity due to the presence of salt becomes significant (demonstrated in our previous work).[4] Here, we show the survival probability plot comparing our evaporation model against that of the simple model. The linear model predicts a maximum S of more than 4 which is highly unlikely for NaCl-water system.

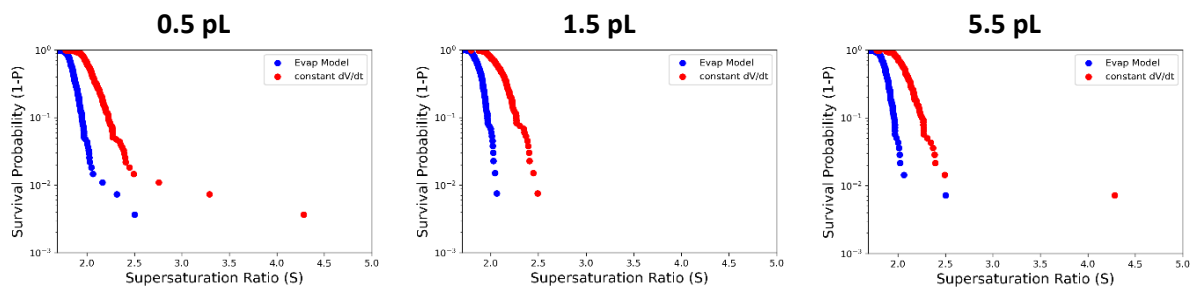

**Figure S3.** Survival probability plot of supersaturation ratios at nucleation calculated using a system-specific evaporation model (blue) compared against that of a simple linear model (red).

## S4.2. Sensitivity with respect to the empirical parameter of the evaporation model

The evaporation model[4] contains an empirical parameter $n_x$ which corresponds to the effective number of neighboring droplets that contribute to the local relative humidity. This parameter $n_x$ was adjusted for each droplet size such that the experimental matching time is accurately reproduced. Here, we show that the resulting distribution of $S_N$ (supersaturation at nucleation) is not sensitive to this empirical parameter. Therefore, the estimated CNT parameters are not sensitive to the adjustment of $n_x$.

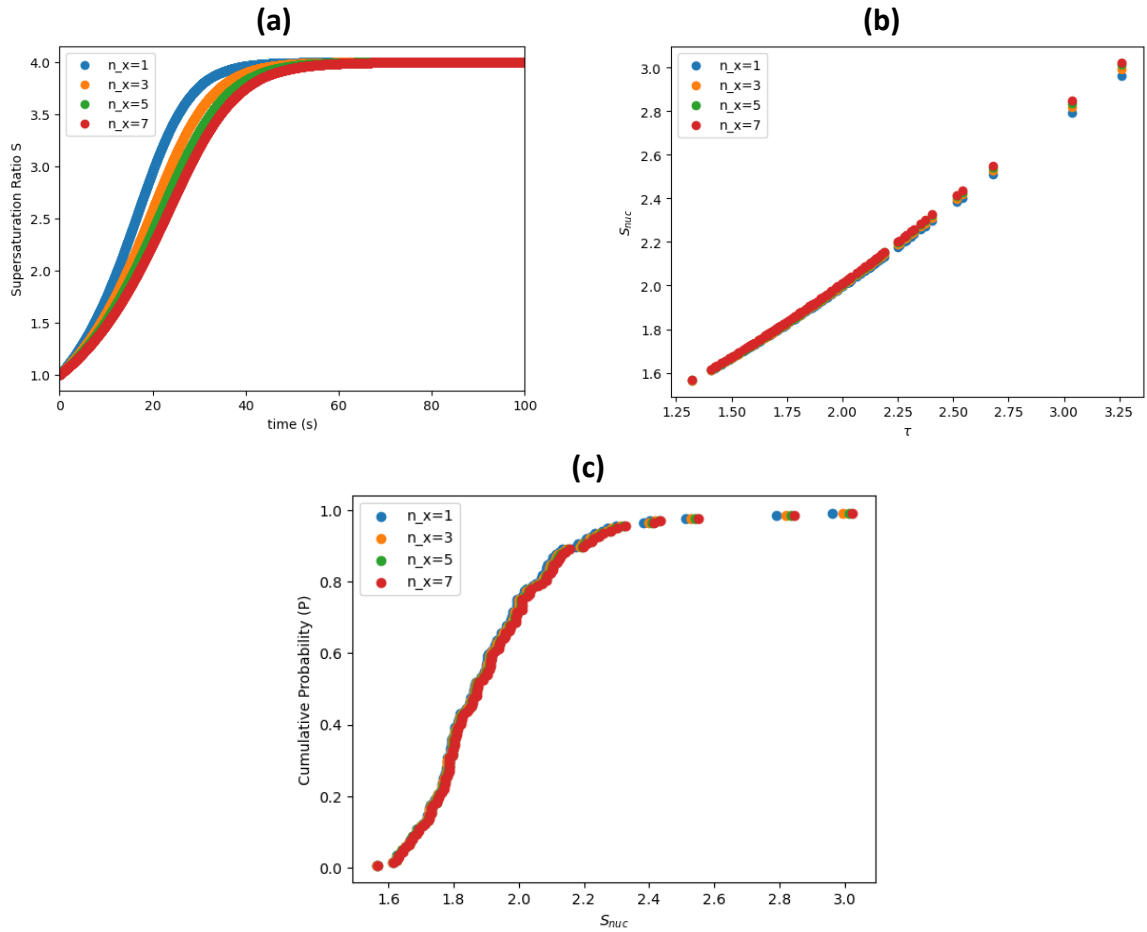

**Figure S4**. Impact of different values of empirical parameter $n_x$ on (a) Supersaturation vs time (b) supersaturation ratio $S_{NUC}$ vs normalized time τ (c) probability distribution of $S_{NUC}$

## S5. Data Processing and Curve fitting method

From the microscopy images, we analyzed the standard deviation of the gray-level pixels σ of the region surrounding a microdroplet. The plot of σ vs time allows the extraction of characteristic times $t_{sat}$ (saturation time), $t_{match}$ (matching time), and $t_{nuc}$ (nucleation time). From these characteristic times, we can calculate a dimensionless nucleation time

$$\tau = \frac{t_{nuc} - t_{sat}}{t_{match} - t_{sat}}$$

More details of this procedure (as well as the statistical treatment of outliers) is in Ref.[5]

To obtain the supersaturation ratio $S$ at any $\tau$, we employed a tailored evaporation model.[4]

The evaporation model contains an empirical parameter $n_x$ which can be adjusted such that the theoretical matching time $\tilde{t}_{match}$ reproduces the experimental median matching time $\bar{t}_{match}$ after saturation (results are not sensitive to $n_x$ adjustments as shown in S4.2).

The evaporation model then allows the calculation of supersaturation ratio $S(t)$ and volume $V(t)$ as a function of time $t$. Note that the time variable $t$ in the evaporation model is referenced with respect to $t_{sat}$ (i.e., $t = 0$ at $S = 1$). Consequently, the correspondence between the $t$-scale and $\tau$-scale is given by

$$\tau = \frac{t}{\tilde{t}_{match}}$$

This correspondence allows as to calculate $S(t)$ and $V(t)$ at any given $\tau$ which are needed in the evaluation of the inhomogeneous Poisson equation coupled with classical nucleation theory as follows:

$$P(t) = 1 - \exp\left[-\int_{t_{sat}}^{t_{nuc}} J(t)V(t)dt\right]$$

where

$$J(t) = A \exp\left[-\frac{16\pi}{3}\frac{\gamma_{eff}^3}{\rho_s^2(k_bT)^3 \ln^2 S(t)}\right]$$

Note that $P(t)$ is coupled simultaneously with $J(t)$ to find the values of $A$ and $\gamma_{eff}$ that minimize the squared residuals between the experimental and modeled $P(t)$, similar to that of Goh et al[2]. This means that the nucleation rate $J$ is accessible after the fitting. For the curve fitting, we used the least-squares method 'leastsq' as implemented in the LMFIT[6] module.

## S6. Effect of Solubility on Thermodynamic Confinement

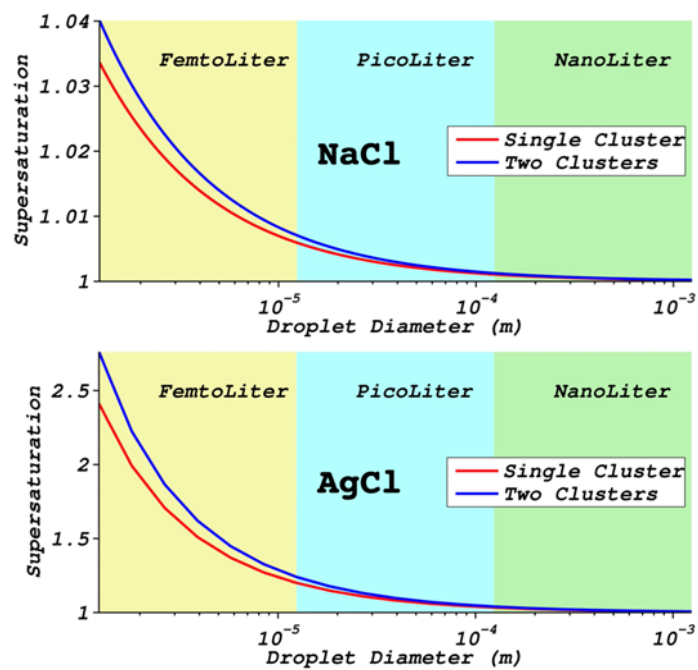

**Figure S5.** Effective supersaturation upon cluster formation as a function of droplet diameter for model compounds NaCl (high solubility) and AgCl (low solubility). Due to the lower solubility of AgCl, confinement-induced depletion of supersaturation needed to form a single critical-cluster is more dramatic effect in AgCl than NaCl. (Adapted from Ref[7]).